\documentclass[lettersize,journal]{IEEEtran}
\usepackage{amsmath,amsfonts}
\usepackage{algorithmic}
\usepackage{algorithm}
\usepackage{array}
\usepackage[caption=false,font=normalsize,labelfont=sf,textfont=sf]{subfig}
\usepackage{textcomp}
\usepackage{stfloats}
\usepackage{url}
\usepackage{marvosym}
\usepackage{verbatim}
\usepackage{graphicx}
\usepackage{cite}

\begin{document}

\title{NAFRSSR: a Lightweight Recursive Network for Efficient Stereo Image Super-Resolution}

\author{
        Yihong~Chen,
        Zhen Fan,
	  Shuai Dong, 
        Zhiwei Chen, 
        Wenjie Li, 
        Minghui Qin,
        Min Zeng,
        Xubing Lu,
        Guofu Zhou,
        Xingsen Gao,
        Jun-Ming Liu

\thanks{

This work was This work was supported in part by the Science and Technology Projects in Guangzhou (202201000008), National Natural Science Foundation of China (Nos. 92163210 and 52172143), and Science and Technology Program of GuangZhou (No. 2019050001). (Corresponding author: Zhen Fan.))

Yihong Chen, Zhen Fan, Shuai Dong, Zhiwei Chen, Wenjie Li,  Minghui Qin, Min Zeng, Xubing Lu and Xingsen Gao are with the Institute for Advanced Materials and Guangdong Provincial Key Laboratory of Optical Information Materials and Technology, South China Academy of Advanced Optoelectronics, South China Normal University, 510006 Guangzhou, China. (e-mail: c1663288225@163.com, fanzhen@m.scnu.edu.cn, ds1426481894@gmail.com, zw\textunderscore chen\textunderscore flyway123@163.com, 1378803662@qq.com,  qinmh@scnu.edu.cn, zengmin@scnu.edu.cn, luxubing@m.scnu.edu.cn, xingsengao@scnu.edu.cn).

 Guofu Zhou is with the National Center for International Research on Green Optoelectronics, South China Normal University, 510006 Guangzhou, China. (e-mail:guofu.zhou@m.scnu.edu.cn).

Jun-Ming Liu is with Laboratory of Solid State Microstructures and Innovation Center of Advanced Microstructures, Nanjing University, 210093 Nanjing, China. (e-mail:liujm@nju.edu.cn).

}

}


\maketitle

\begin{abstract}
Stereo image super-resolution (SR) refers to the reconstruction of a high-resolution (HR) image from a pair of low-resolution (LR) images as typically captured by a dual-camera device. To enhance the quality of SR images, most previous studies focused on increasing the number and size of feature maps and introducing complex and computationally intensive structures, resulting in models with high computational complexity. Here, we propose a simple yet efficient stereo image SR model called NAFRSSR, which is modified from the previous state-of-the-art model NAFSSR by introducing recursive connections and lightweighting the constituent modules. Our NAFRSSR model is composed of nonlinear activation free and group convolution-based blocks (NAFGCBlocks) and depth-separated stereo cross attention modules (DSSCAMs). The NAFGCBlock improves feature extraction and reduces number of parameters by removing the simple channel attention mechanism from NAFBlock and using group convolution. The DSSCAM enhances feature fusion and reduces number of parameters by replacing 1$\times$1 pointwise convolution in SCAM with weight-shared 3$\times$3 depthwise convolution. Besides, we propose to incorporate trainable edge detection operator into NAFRSSR to further improve the model performance. Four variants of NAFRSSR with different sizes, namely, NAFRSSR-Mobile (NAFRSSR-M), NAFRSSR-Tiny (NAFRSSR-T), NAFRSSR-Super (NAFRSSR-S) and NAFRSSR-Base (NAFRSSR-B) are designed, and they all exhibit fewer parameters, higher PSNR/SSIM, and faster speed than the previous state-of-the-art models. In particular, to the best of our knowledge, NAFRSSR-M is the lightest (0.28M parameters) and fastest (50 ms inference time) model achieving an average PSNR/SSIM as high as 24.657 dB/0.7622 on the benchmark datasets. Codes and models will be released at \url{https://github.com/JNUChenYiHong/NAFRSSR}.
This paper was finished in March 2023, so no subsequent notable works are included. Declaration made..
\end{abstract}

\begin{IEEEkeywords}
Stereo image super-resolution, Recursive connection, Lightweight network, Edge detection.
\end{IEEEkeywords}

\section{Introduction}
\subsection{Image super-resolution}
\IEEEPARstart{C}{onventionally}, obtaining high-quality images requires the use of high-resolution cameras (such as 4K/8K cameras), which are however bulky, power-hungry, and expensive. To circumvent this issue, researchers have proposed an image super-resolution (SR) technology, through which the resolution and quality of images can be improved. Specifically, image SR refers to the reconstruction of a high-resolution (HR) image from a low-resolution (LR) counterpart given an up-scaling factor using a specific method (e.g. traditional interpolation algorithms and deep learning). The reconstructed image is typically called an SR image. In recent years, deep learning-based methods have attracted wide attention in the SR field because of their good reconstruction performance. Different from the traditional interpolation-based methods, the deep learning-based methods use a large number of image samples as priors to constrain the reconstruction process. Such image samples contain rich textural information, which is encoded into the parameters of the deep learning models. These models can then reconstruct the fine image details by decoding, thus improving the subjective and objective quality of the SR images.  
\subsection{Stereo image super-resolution}

Nowadays binocular imaging has been widely used in mobile phone cameras, remote sensing, reconnaissance and surveillance, intelligent robots, and other fields. Along with the development of binocular imaging, stereo image SR also emerges, which aims at reconstructing SR images from a pair of stereo images on the left and right views captured by a binocular camera system. To obtain high-quality SR images, the key is to make full use of the complementary feature information provided by the stereo image pair (i.e., the cross-view information\cite{SRRes+SAM}).Most previous works on stereo image SR were deep learning-based. They mainly used attention mechanisms to exploit the cross-view information, such as parallax-attention mechanism\cite{PASSRnet}, stereo attention module\cite{SRRes+SAM}, disparity attention losses\cite{StereoIRN}, and bi-directional parallax attention\cite{iPASSR}. Although these deep learning models demonstrated competitive performance in stereo image SR tasks, they were typically complex and time consuming. Recently, NAFSSR\cite{NAFSSR} was developed by using a simple yet efficient image restoration model, NAFNet\cite{NAFNet}, for the intra-view feature extraction, and adding cross attention modules to fuse left- and right-view features. NAFSSR achieved the state-of-the-art performance on the KITTI 2012\cite{KITTI2012}, KITTI 2015\cite{KITTI2015}, Middlebury\cite{Middlebury}, and Flickr1024\cite{flickr1024} datasets. Moreover, it contained a smaller number of parameters than the previous state-of-the-art model, SSRDE-FNet\cite{SSRDE-FNet} (0.46M versus 2.24M), and exhibited a 5.11$\times$ speedup. 
However, there is still room for improvement in the performance of NAFSSR, as its building blocks, i.e., NAFBlock and SCAM (for extracting intra-view and cross-view features, respectively), can be further optimized. Specifically, the channel attention mechanism in NAFBlock can only work when the number of channels is large, but in such case the number of parameters related to the channel attention mechanism is quite large. On the other hand, SCAM has a complex structure and a large number of unnecessary parameters that are irrelevant to the adequate extraction of cross-view information. 

\subsection{Contribution}
Motivated by the above facts, we modify NAFSSR by introducing recursive connections and lightweighting its building blocks (NAFBlock and SCAM), resulting in a lightweight efficient stereo image SR model called NAFRSSR. Compared with the previous state-of-the-art models, NAFRSSR exhibits fewer parameters, higher PSNR/SSIM, faster inference speed, and better visual effects. The main contributions of our work can be summarized as follows: 

\begin{enumerate}
  \item We construct a weight-sharing depth-separated stereo cross attention module (DSSCAM) by replacing the pointwise (PW) convolution in SCAM with a weight-sharing 3$\times$3 depthwise (DW) convolution. Compared with SCAM, DSSCAM can better fuse complementary feature information between stereo images with fewer parameters.
  \item We construct a nonlinear activation free and group convolution-based block (NAFGCBlock) by removing the channel attention mechanism in NAFBlock and introducing group convolution. Besides, recursive connections are applied to certain NAFGCBlocks. With these modifications, the intra-view feature extraction capability of the stack of NAFGCBlocks is enhanced while the number and size of NAFGCBlocks are reduced. 
  \item We incorporate a trainable edge detection operator into the last layer of our model to further recover the edge (i.e. high-frequency) feature information of an SR image, thereby enhancing its quality. 
 \item Based on the DSSCAM, NAFGCBlock and trainable edge detection operator, we propose NAFRSSR for efficient stereo image SR. Compared with the previous state-of-the-art models SwinFIR\cite{SwinFIR} and NAFSSR, NAFRSSR demonstrates lower parameters, higher average PSNR/SSIM, faster inference speed, and better visual effects.
  
\end{enumerate}


\section{Related Works}
\subsection{ Single image super-resolution models}
Nowadays, there are mainly two types of methods to achieve single image SR: traditional interpolation-based methods (such as BICUBIC) and deep learning-based methods (such as SRCNN\cite{SRCNN}). Many studies have demonstrated that the deep learning-based methods can achieve better reconstruction results than the traditional interpolation-based methods on image SR tasks. In 2014, Dong et al. introduced the deep convolutional neural network into the single image SR for the first time. Their proposed SRCNN directly learned an end-to-end mapping between the low- and high-resolution images and achieved better performance than traditional algorithms (27.50 dB PSNR on the 4$\times$ SR task of the Set14 dataset). Since then, many deep learning-based image SR models have been proposed, such as ESPCN\cite{ESPCN}, VDSR\cite{VDSR}, WDSR\cite{WDSR}, DRCN\cite{DRCN}, RCAN\cite{RCAN}, SwinIR\cite{SwinIR}, and SwinFIR\cite{SwinFIR}. Particularly, Liang et al. proposed SwinIR based on Swin Transformer and achieved PSNR/SSIM of 32.22 dB/0.9273 on the Manga109 dataset. Furthermore, Zhang et al. modified SwinIR by replacing the convolution layer of the deep feature extraction module with Fast Fourier Convolution (FFC) which can extract the global information in the frequency branch. The resulting SwinFIR model achieved the state-of-the-art performance (32.83 dB PSNR on the Manga109 dataset).

\subsection{Stereo image super-resolution models}
Different from the single image SR, the stereo image SR uses a pair of LR images to reconstruct an HR image. In order to learn the feature mapping relationship between the stereoscopic LR images and the HR image, many deep learning-based models have been developed. 

In 2018, the first stereo image SR model called StereoSR\cite{StereoSR} was proposed, which utilized two cascaded sub-networks for joint training in the directions of brightness and chrominance, achieving an average PSNR/SSIM value of 22.569/0.6783 on all test sets. In 2019, PASSRNet\cite{PASSRnet}introduced the Parallax Attention Mechanism Module (PAM), which can fuse complementary information along the parallax direction of stereo images without the limit of the parallax size. Compared with StereoSR, PASSRNet  was more flexible and stable. In 2020, Ying et al. proposed the Stereo Attention Module (SAM)\cite{SRRes+SAM}, and inserted SAMs into the existing stereo image SR models (such as SRCNN\cite{SRCNN}, VDSR\cite{VDSR}, LapSRN\cite{LapSRN}, SRDenseNet\cite{SRDenseNet}, SRResNet\cite{SRResNet}), yielding good results. For example, SRRes+SAM achieved 23.27 dB PSNR on the Flickr1024\cite{flickr1024} dataset. In the same year, StereoIRN\cite{StereoIRN} was proposed, which interpreted the inherent correspondence between pixels of stereo images and utilized the intersection features between these stereo images for image reconstruction. Additionally, StereoIRN utilized a Feature Modulation Dense Block (FMDB) to incorporate disparity prior information into the network and achieved a PSNR of 29.83 dB on all left images in the Middlebury\cite{Middlebury} dataset. SPAMNet\cite{SPAMNet},as proposed by Song et al. In 2020, employed self-attention and parallel attention mechanism (SPAM)to simultaneously extract image features, and fuse complementary feature information of the image pair. In 2021, IPASSR\cite{iPASSR} introduced a bidirectional parallax attention module (biPAM) and an inline occlusion handling scheme, which can efficiently process occluded areas and can process images of left and right views simultaneously. SSRDE-FNet\cite{SSRDE-FNet} achieved 23.59 dB PSNR on the Flickr1024 dataset by using cross-view information to improve the SR image and high-resolution features for better disparity estimation. In 2022, Chu et al. proposed NAFSSR\cite{NAFSSR}, which added a channel attention mechanism and a stereo cross-attention module to the single image SR model NAFNet\cite{NAFNet}. NAFSSR achieved +0.05 dB higher PSNR than SSRDE-FNet, while being five times faster and using only 20.5\% of the parameters. 

Although the above deep learning-based models have achieved good performance in stereo image SR tasks, they typically still have complex structures and large numbers of parameters. This motivates us to develop a simple yet efficient stereo image SR model, as described in detail in the next section.

\section{Method}
In this chapter, we will describe in detail the proposed lightweight efficient stereo image SR model NAFRSSR. NAFRSSR mainly consists of two weight-sharing network branches (stacked by NAFGCBlocks) with DSSCAMs connected between them and a trainable edge detection operator inserted at the branch end. The overall architecture and constituent modules of NAFRSSR will be described in detail as follows. 

\subsection{Overall network architecture}
\label{sec:Overall_architecture}
\begin{figure*}[!t]
  \centering
  \includegraphics[width=7in, keepaspectratio]{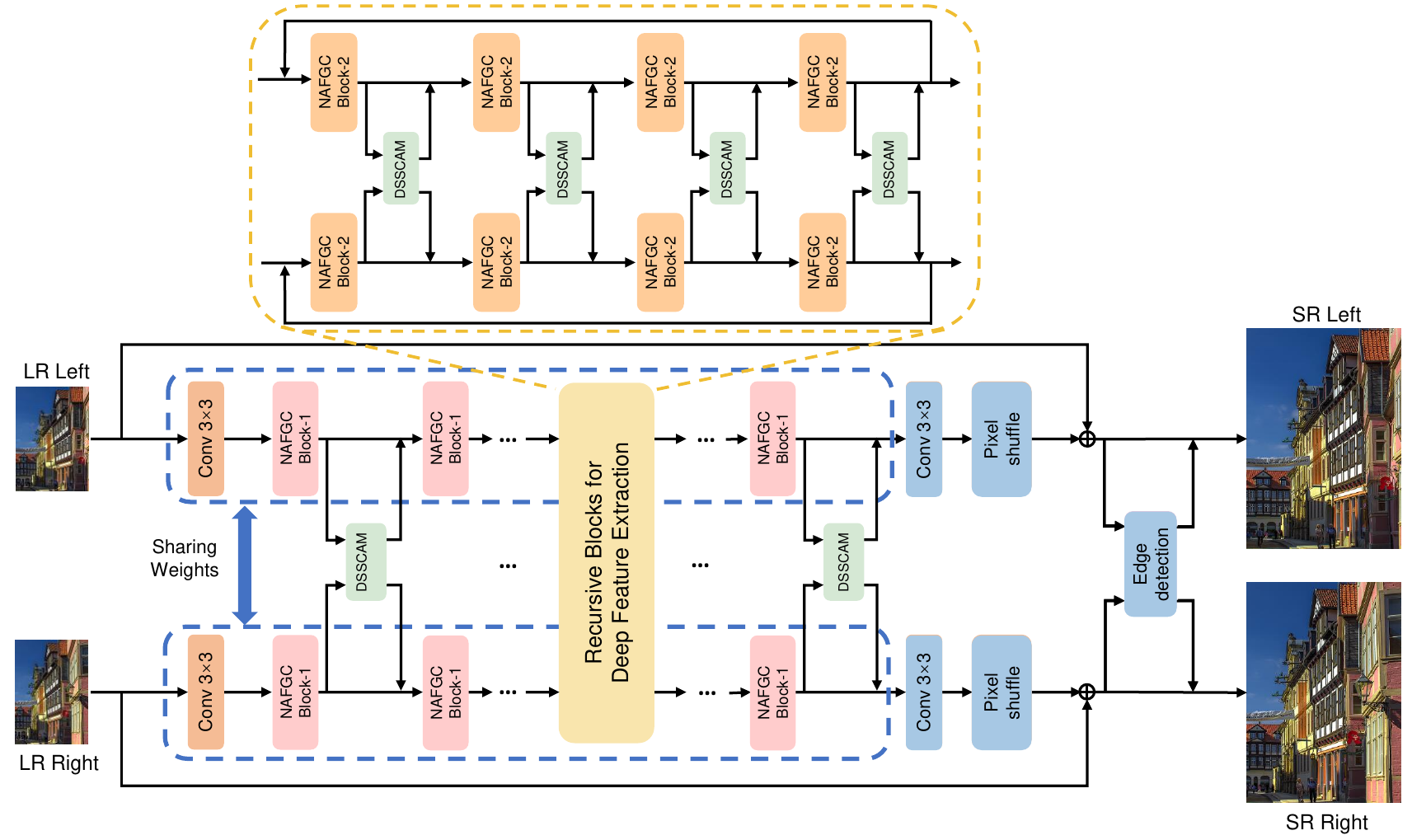}
  \caption{The overall architecture of NAFRSSR. NAFGCBlock represents nonlinear activation free and group convolution-based block. DSSCAM represents weight-sharing depth-separated stereo cross attention module.}
  \label{fig1}
\end{figure*}

The overall architecture of the NAFRSSR is shown in Fig. \ref{fig1}, which is similar to that of NAFSSR. It mainly consists of two weight-sharing network branches (stacked by NAFGCBlocks) to process the left- and right-view LR images, respectively, and DSSCAMs are added between the two weight-sharing network branches to perform the cross-view feature fusion. According to the dataflow, NAFRSSR can be divided into three sequential parts: feature mapping layer, deep feature extraction layer, and upsampling layer. After passing through the three layers, the left- and right-view LR images will be upsampled, and then combined with their corresponding SR images produced by the bicubic interpolation algorithm. The combined SR images are then subjected to edge detection to generate the final output SR images.

Below we will introduce the three layers of NAFRSSR in more detail.
\begin{enumerate}
    \item{Feature mapping layer: It is a 3$\times$3 convolution layer, which is responsible for preliminarily extracting the features of a pair of stereo LR images and mapping the image features to a high-dimensional space.}
    
    \item{Deep feature extraction layer: It consists of N NAFGCBlocks that are stacked in series, and N DSSCAMs, each of which is placed after each NAFGCBlock and is bridged between the two network branches. Note that recursive connections are applied to some NAFGCBlocks, i.e., NAFGCBlock-2s, while the rest, i.e., NAFGCBlock-1s, are free of the recursive connections. The left- and right-view images’ feature maps output by NAFGCBlock$_{n-1}$ will be input to the subsequent DSSCAM$_{n-1}$ for the cross-view feature fusion. Then the sum of the outputs of NAFGCBlock$_{n-1}$ and DSSCAM$_{n-1}$ are sent to the next NAFGCBlock$_{n}$ as the input. This layer is responsible for the deep feature extraction and feature fusion of the input stereo image pairs, and it finally outputs high-dimensional feature maps as the inputs of the following upsampling layer.}
    
    \item{Upsampling layer: It consists of a 3$\times$3 convolution layer, a pixel shuffle layer\cite{pixelShuffle} and a trainable edge detection operator. It integrates the feature maps output from the deep feature extraction layer, and then upsamples them to reconstruct the SR image.}
    
\end{enumerate}

\subsection{Depth-separated stereo cross attention module}

\begin{figure}[!t]
  \centering
  \includegraphics[width=3.5in, keepaspectratio]{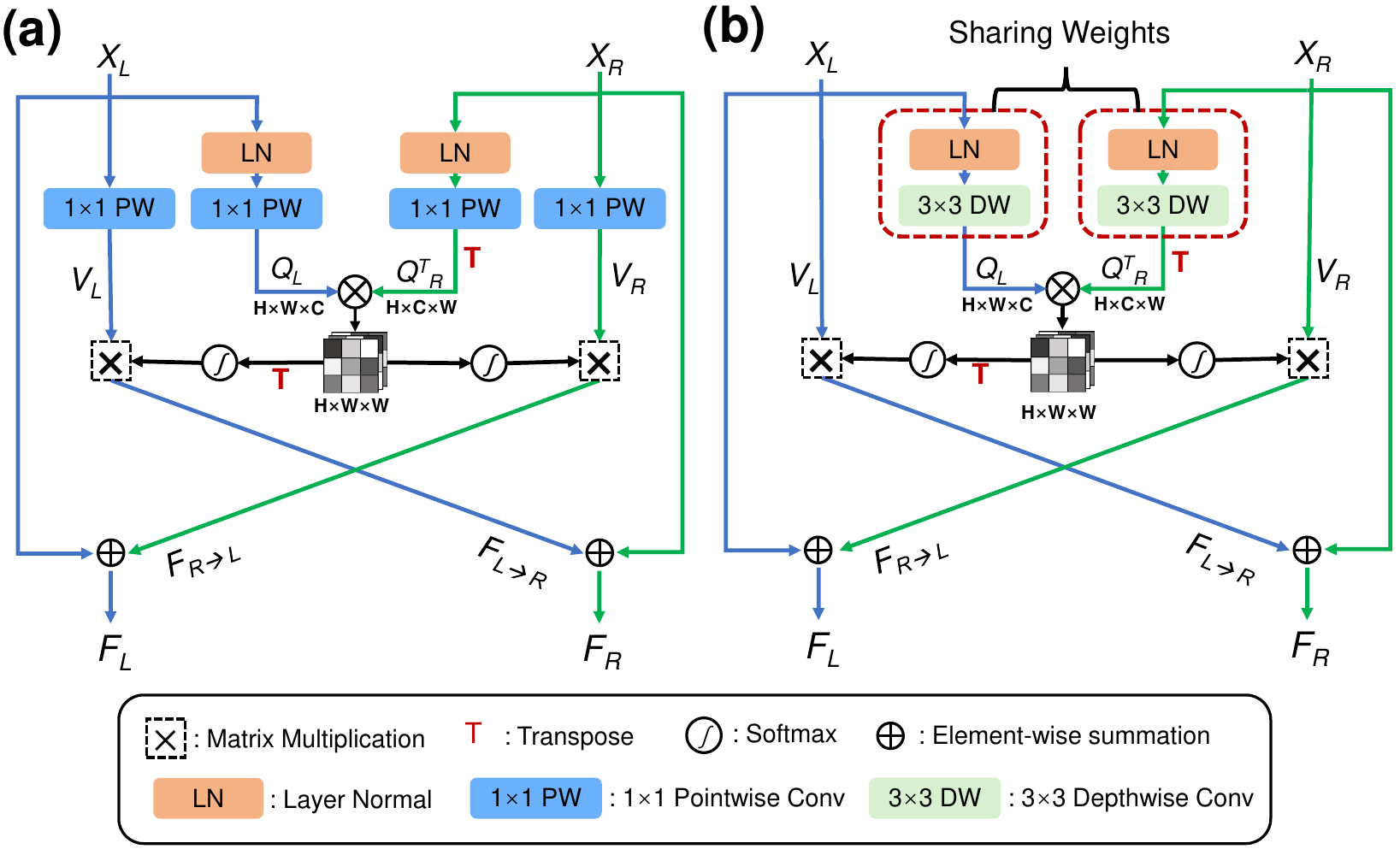}
  \caption{Comparison between (a) stereo cross attention module (SCAM) of NAFSSR and (b) depth-separated stereo cross attention module (DSSCAM) of NAFRSSR. }
  \label{fig2}
\end{figure}

Fig. \ref{fig2}a and b compares the structures of SCAM of NAFSSR and DSSCAM of our NAFRSSR. Both SCAM and DSSCAM have the same function, i.e., fusing the features of a pair of stereo images and extract the complementary feature information of them. Specifically, both SCAM and DSSCAM take the feature maps of the left- and right-view images, i.e., $X_L$ and $X_R \in R^{H \times W \times C}$, respectively, as the inputs, where H, W, and C represent the height, width, and channel of feature maps. $X_L$ and $X_R$ are used to extract the interacted cross-view information $F_{R \to L}$ and $F_{L \to R}$ based on the cross attention:

\begin{equation}
\label{eq.1}
F_{R \to L}={Softmax}\Big(Q_L Q_{R}^{T}/\sqrt{C}\Big)V_R
\end{equation} 
\begin{equation}
\label{eq.2}
F_{L \to R}={Softmax}\Big(Q_R Q_{L}^{T}/\sqrt{C}\Big)V_L
\end{equation} 

where $Q_L$ ($Q_R$) and $V_L$ ($V_R$) stand for the query and value matrices projected by $X_L$ ($X_R$). Note that the projections from $X_L$ ($X_R$) to $Q_L$ ($Q_R$) and $V_L$ ($V_R$) are different in SCAM and DSSCAM (to be described later). In addition, $Q_R$ and $Q_L$ are also used as the key matrices for $X_L$ and $X_R$, respectively (see Eq. \ref{eq.1} and \ref{eq.2}), based on the fact that stereo images are highly symmetric under the epipolar constraint\cite{iPASSR}. 

After obtaining  $F_{R \to L}$ and $F_{L \to R}$, they are fused with $X_L$ and $X_R$, respectively, by using element-wise addition: 
\begin{equation}
\label{eq.3}
F_L= \gamma_{L} F_{R \to L} + X_L
\end{equation} 
\begin{equation}
\label{eq.4}
F_R= \gamma_{R} F_{L \to R} + X_R
\end{equation} 

where $\gamma_{L}$ and $\gamma_{R}$ are trainable weight parameters, $F_L$ and $F_R$ are the final outputs of the SCAM or DSSCAM. 
As mentioned above, how $Q_L/Q_R$ and  $V_L/V_R$  are projected from $X_L/X_R$ are different in SCAM and DSSCAM. In SCAM (see Fig. \ref{fig2}a), $Q_L/Q_R$ are the outputs of $X_L/X_R$ after layer normalization (LN)\cite{LN} and PW convolution\cite{mobilenet}, while $V_L/V_R$ are the outputs of $X_L/X_R$ after different PW convolutions. The $V_L/V_R$ and $Q_L/Q_R$ in SCAM can be written as follows:

\begin{equation}
\label{eq.5}
Q_L= PW\Big(LN(X_L)\Big)
\end{equation} 
\begin{equation}
\label{eq.6}
Q_R= PW\Big(LN(X_R)\Big)
\end{equation} 
\begin{equation}
\label{eq.7}
V_L= PW\Big(X_L\Big)
\end{equation} 
\begin{equation}
\label{eq.8}
V_R= PW\Big(X_R\Big)
\end{equation} 

By contrast, in DSSCAM (see Fig. \ref{fig2}b), $Q_L/Q_R$ are the outputs of $X_L/X_R$  after weight-sharing LN and 3$\times$3 depthwise (DW) convolution\cite{mobilenet}, while $V_L/V_R$ are equal to $X_L/X_R$ . The $Q_L/Q_R$ and $V_L/V_R$ in DSSCAM can be written as follows: 
\begin{equation}
\label{eq.9}
Q_L= DW\Big(LN(X_L)\Big)
\end{equation} 
\begin{equation}
\label{eq.10}
Q_R= DW\Big(LN(X_R)\Big)
\end{equation} 
\begin{equation}
\label{eq.11}
V_L= X_L
\end{equation} 
\begin{equation}
\label{eq.12}
V_R= X_R
\end{equation} 
Compared with SCAM, DSSCAM uses two weight-sharing 3$\times$3 DW convolutions instead of two independent PW convolutions for calculating $Q_L/Q_R$, which can result in a larger receptive field and fewer parameters. In addition, the weight sharing strategy is employed for both the LN and DW convolutions in DSSCAM, which not only reduces the number of parameters, but also better encodes the correlated feature information of stereo images. On the other hand, for calculating $V_L/V_R$ , the PW convolution is removed in DSSCAM. This not only simplifies the calculation, but also better retains the original feature information of the left- and right-view images.

\subsection{Nonlinear activation free and group convolution-based block}
\label{sec:NAFGCBlock_stc}
\begin{figure}[!t]
  \centering
  \includegraphics[width=3.4in, keepaspectratio]{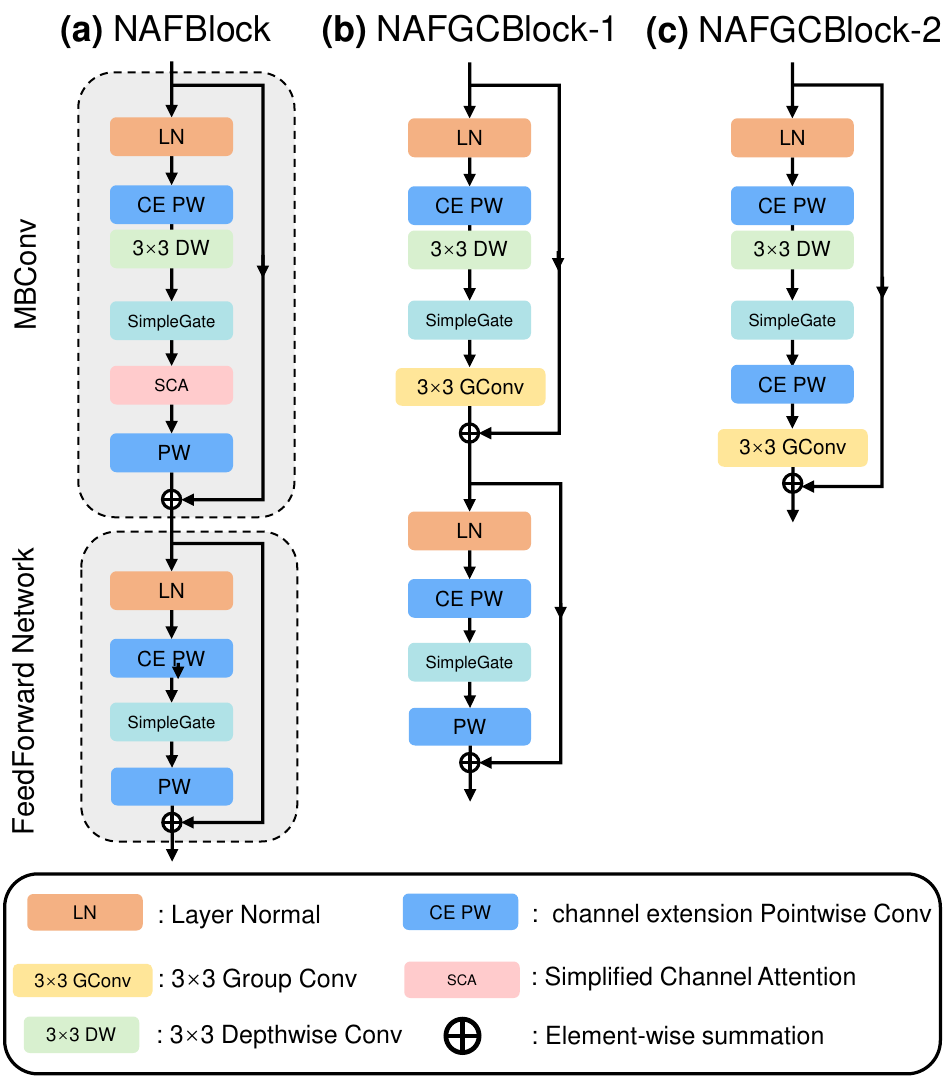}
  \caption{(a) Nonlinear activation free block (NAFBlock) of NAFSSR. Two types of nonlinear activation free and group convolution-based blocks (NAFGCBlocks): (b) NAFGCBlock-1 and (c) NAFGCBlock-2.}
  \label{fig3}
\end{figure}

\begin{figure}[!t]
  \centering
  \includegraphics[width=3in, keepaspectratio]{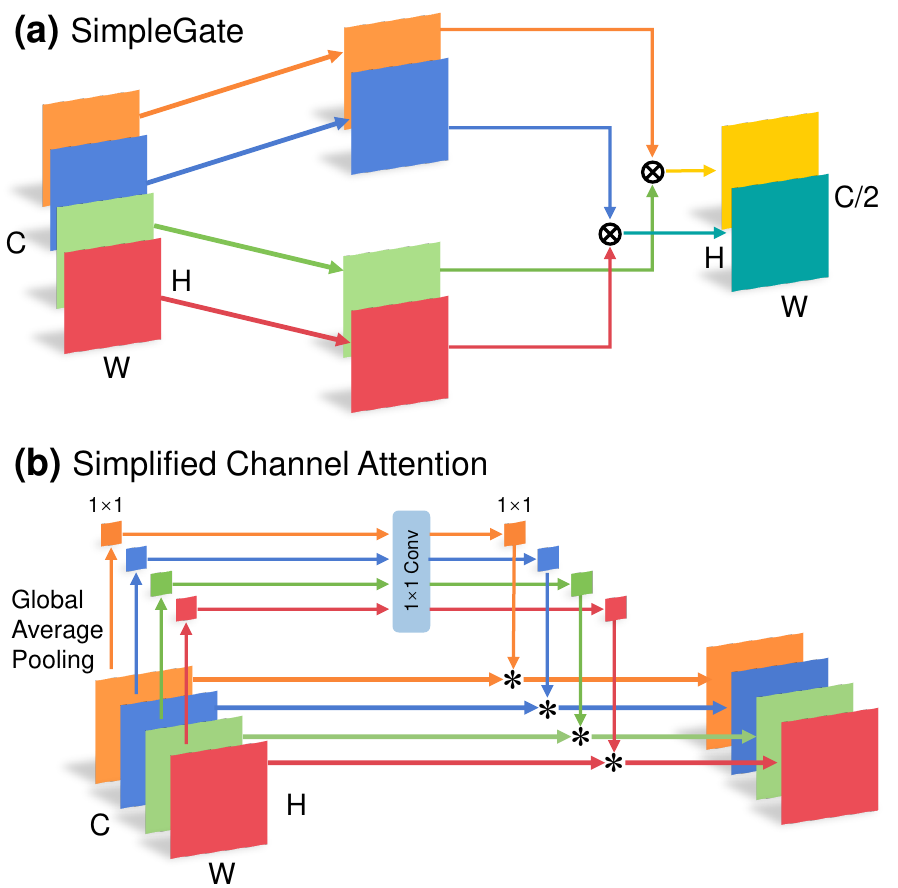}
  \caption{Schematic diagrams of (a) SimpleGate (SG) and (b) Simplified channel attention (SCA).}
  \label{fig4}
\end{figure}

Fig. \ref{fig3}a shows the feature extraction module NAFblock of NAFSSR. It is composed of a mobile convolution module (MBConv)\cite{mobilenetV3}and a feedforward network (FFN)\cite{Attention}. MBConv consists of LN, PW convolution, channel extension PW convolution (CEPW), DW convolution, SimpleGate (SG)\cite{NAFNet} and simplified channel attention (SCA)\cite{NAFNet}, while FFN consists of LN, CEPW, PW convolution, and SG. 
More specifically, CEPW represents the use of PW convolution to expand the number of channels of the feature maps by s times. SG is a nonlinear activation function (see Fig. \ref{fig4}a),which can be written as follows: 
\begin{equation}
\label{eq.13}
SimpleGate(X)= X_1 \otimes X_2
\end{equation} 

where $X \in R^{H \times W \times C}$  represents the feature map input to SG, $X_1$  and $X_1$ are the two parts of $X$ as split along the $C$ dimension, and $\otimes$ represents the element-wise multiplication. SCA refers to a simple channel attention mechanism (see Fig. \ref{fig4}b), which comprises global average pooling and a point-wise (PW) convolution using a trainable weight vector $W\in R^{1 \times 1 \times C}$, where * denotes a channel-wise product operation: 
\begin{equation}
\label{eq.14}
SCA(X)= X*\Big(W AveragePool(X)\Big)
\end{equation} 

Our proposed NAFGCBlock-1 and NAFGCBlock-2 are modified from NAFBlock, as shown in Fig. \ref{fig3}b and c, respectively. The modification details are described as follows. First, SCA is removed in both NAFGCBlock-1 and 2 because our experiment reveals that SCA is unnecessary (to be shown in Section \ref{sec:NAFGCBlock}). Second, NAFGCBlock-1 replaces the last PW convolution of MBConv with a 3$\times$3 group convolution (GConv). In this way, the parameters of NAFGCBlock-1 is reduced compared to the NAFBlock of NAFRSSR, while their feature extraction capability is improved. In addition, NAFGCBlock-2 is further simplified, consisting of LN, CEPW, 3$\times$3 DW, SG, CEPW, and 3$\times$3 GConv only. 
Note that NAFGCBlock-1 and NAFGCBlock-2 are designed for different purposes: NAFGCBlock-1 is designed for pursuing high performance, while NAFGCBlock-2 is designed to accommodate the recursive structure to reduce the number of model parameters. 

\subsection{Edge detection and super-resolution}
\label{sec:Edge_method}
Edge detection is commonly used to extract contour information (i.e. high frequency information features) from images. In this paper, we propose to apply edge detection to the stereo image SR task.
\begin{figure}[!t]
  \centering
  \includegraphics[width=3.4in, keepaspectratio]{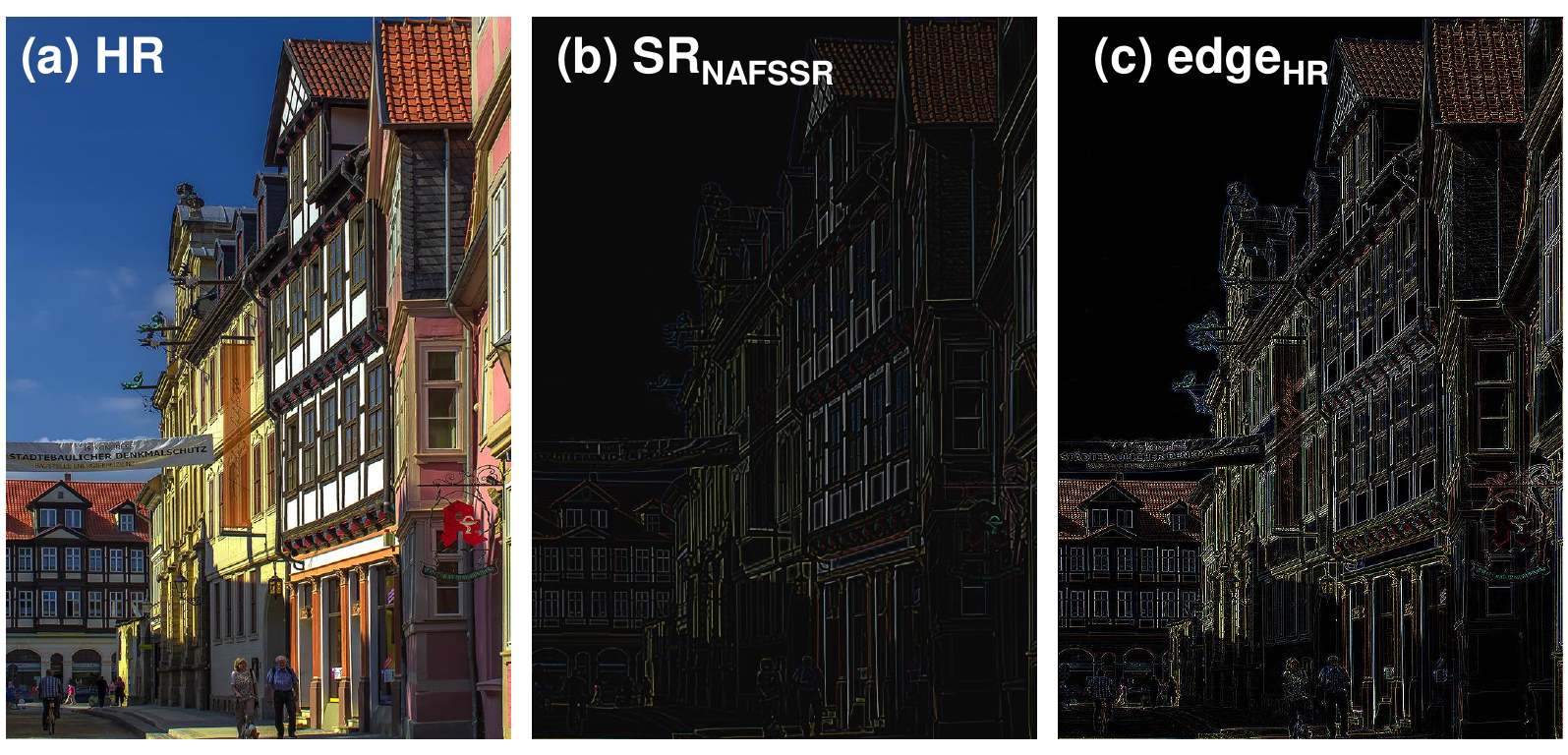}
  \caption{(a) The input HR image, (b) $SR_{NAFSSR}$ is the feature map output by the NAFSSR model and (c) $edge_{HR}$ is the result of the edge detection of the HR image. }
  \label{fig5}
\end{figure}

This idea comes when inspecting the feature map output by the last layer of NAFSSR (i.e. $SR_{NAFSSR}$). Fig. \ref{fig5}a and b show the input HR image and correspongding $SR_{NAFSSR}$ image. After $SR_{NAFSSR}$ is obtained, it is added with the SR image generated by the bicubic interpolation algorithm to yield the final output, i.e., $SR_{output} = SR_{NAFSSR} + SR_{bicubic}$. It is interesting to note that the $SR_{NAFSSR}$ image displays similar features to the edge detection result of the input HR image (see comparison between Fig. \ref{fig5}b and c. One may thus conjecture that the edge information is very important to reconstruct a good SR image.
We therefore propose to apply an edge detection operator to (SR$_{NAFSSR}$ + SR$_{bicubic}$), and then sum the edge detection result, SR$_{NAFSSR}$, and SR$_{bicubic}$ to obtain the final output, as expressed by:
\begin{equation}
\label{eq.15}
SR_{output} = SR_{NAFRSSR} + SR_{bicubic} + x \cdot SR_{edge}
\end{equation} 
\begin{equation}
\label{eq.16}
SR_{edge} = edge\Big(SR_{NAFRSSR} + SR_{bicubic}\Big)
\end{equation} 
where $x$ represents trainable weight parameters, and the function $edge$  represents the edge detection calculation. The size of the edge detection operator is 3$\times$3, and its parameters are trainable. The addition of the edge detection operator can improve the image edge information reconstruction performance of the model with little increase in the number of model parameters.

\subsection{Network size}
For our NAFRSSR, in order to achieve a tradeoff between the performance and inference time, the numbers of the feature map channels and NAFGCBlocks can be varied. We thus design 4 different types of NAFRSSR models: AFRSSR-Mobile (NAFRSSR-M), NAFRSSR-Tiny (NAFRSSR-T), NAFRSSR-Super (NAFRSSR-S) and NAFRSSR-Base (NAFRSSR-B), which may be used for devices with different computing powers (e.g., mobile terminal, PC terminal). Their model sizes are shown in Table~\ref{TABLE1}. 

\begin{table}[htbp]
  \centering
  \caption{Architectures and parameters of four NAFRSSR variants}
  \begin{tabular}{c  c  c  c}
    \hline
    Models & Params & Channels & Blocks\\
    \hline
    NAFRSSR-M & 0.28M & 64 & 16\\
    
    NAFRSSR-T & 0.34M & 72 & 16\\
    
    NAFRSSR-S & 1.35M & 80 & 34\\
    
    NAFRSSR-B & 5.78M & 120 & 70\\
    \hline
    
  \end{tabular}
  \label{TABLE1}
\end{table}

\section{Experimental}
\subsection{Experimental settings}
\subsubsection{Hardware and software systems}
The hardware and software systems used for our experiments are shown in Table~\ref{TABLE2} and Table~\ref{TABLE3}, respectively.

\begin{table}[htbp]
  \centering
  \caption{Hardware system}
  \begin{tabular}{|c | c|}
    \hline
    CPU & 2*E5-2660V3\\
    \hline
    GPU & GeForce RTX 3090 \\
    \hline
    Memory & 2*16GB DDR4\\
    \hline
  \end{tabular}
  \label{TABLE2}
\end{table}

\begin{table}[htbp]
  \centering
  \caption{Software system}
  \begin{tabular}{|c | c|}
    \hline
    OS & Centos 7\\
    \hline
    GPU Driver & 470.86 \\
    \hline
    CUDA & 11.4\\
    \hline
      Python & 3.7 \\
    \hline
    Pytorch & 1.9\\
    \hline
  \end{tabular}
  \label{TABLE3}
\end{table}

\subsubsection{Evaluation Metrics}
Peak signal to noise ratio (PSNR) and structural similarity index (SSIM) are used as the metrics to evaluate the quality of a reconstructed SR image. PSNR measures the general distance between the reconstructed SR image and the corresponding ground truth, while SSIM measures their structural similarity. Higher PSNR and SSIM values typically indicate higher quality of the reconstructed SR image. These metrics are calculated on RGB color space with a pair of stereo images, i.e., (Left + Right) / 2.

\subsubsection{Datasets}
The training set is formed by combining 800 stereo images from Flickr1024\cite{flickr1024} and 60 stereo images from the Middlebury\cite{Middlebury}. The LR images used for training are downsampled from the HR images at a scaling factor of $\times$4 by using the bicubic algorithm. 
For the test, 4 different types of datasets are used: 20 stereo images from KITTI 2012\cite{KITTI2012}, 20 stereo images from KITTI 2015\cite{KITTI2015}, 5 stereo images from Middlebury, and 112 stereo images from Flickr1024. 

\begin{table}[htbp]
  \centering
  \caption{Training datasets and testing datasets}
  \begin{tabular}{|c | c|}
    \hline
    Training sets & Test sets\\
    \hline
     & 112 stereo images from Flickr1024\cite{flickr1024}, \\

    800 stereo images from Flickr1024, & 20 stereo images from KITTI 2012,\\
  
    60 stereo images from Middlebury & 20 stereo images from KITTI 2015,\\
  
    & 5 stereo images from Middlebury \\
    \hline
  \end{tabular}
  \label{TABLE4}
\end{table}

\subsubsection{Training Settings}
All the stereo image SR models, including our NAFRSSR model and other control models, are optimized by using the mean square error (MSE) loss function and the AdamW optimizer, where $\beta_1=0.9$, $\beta_2=0.9$, and weight decay = 0 are set by default. The learning rate is set to be 3 $\times$ $10^{-3}$, which decreases to 1 $\times$ $10^{-7}$ with the cosine annealing strategy\cite{2016sgdr}. To introduce the data augmentation, the full LR images are cropped into small patches (patch size: 30 $\times$ 90; patch number: 49020) with a stride of 20. Then, these small patches are randomly flipped horizontally and vertically to generate the inputs. The 49020 patches with a batch size of 32 are used for training, and the models are trained for a total number of 4 $\times$ $10^5$ iterations.

\subsection{Ablation Study}
\subsubsection{DSSCAM}
The proposed DSSCAM is first compared with the SCAM of NAFSSR to demonstrate its effectiveness. We take the NAFSSR-T model as the baseline, which contains 16 NAFBlocks and 16 SCAMs in total. We then replace all the SCAMs in the NAFSSR-T model with 16 DSSCAMs, resulting a new model called T-DSSCAM. Both the NAFSSR-T and T-DSSCAM models are trained and tested on the same datasets (see Table~\ref{TABLE4}), and their performance is compared.
Table~\ref{TABLE5} shows that T-DSSCAM achieves slightly higher PSNR and SSIM values than NAFSSR-T on various test sets, demonstrating that the capability of T-DSSCAM to interact the cross-view information is slightly better than that of SCAM. Also noteworthy is that the number of parameters in T-DSSCAM is only 68\% of that of NAFSSR-T. It is thus demonstrated DSSCAM is a lighter yet more efficient module than SCAM for the cross-view information fusion. 

\begin{table*}[htbp]
  \centering
  \caption{Quantitative results achieved by NAFSSR-T, T-DSSCAM, T-NoSCA, T-NAFGCBlock-1 and T-edge on the KITTI 2012\cite{KITTI2012}, KITTI 2015\cite{KITTI2015}, Middlebury\cite{Middlebury}, and Flickr1024\cite{flickr1024} datasets. Here, PSNR/SSIM values achieved on a pair of stereo images (i.e., (Left + Right) / 2) are reported. }
  \begin{tabular}{c c c c c c c c}
    \hline
    Method & Parameter & KITTI 2012 & KITTI 2015 & Middlebury & Flickr1024 & Average & $\Delta$PSNR/$\Delta$SSIM \\
    \hline
     NAFSSR-T & 460.66k & 26.7452/0.8099	& 26.5755/0.8154 & 29.3728/0.8404 & 23.6666/0.7390 & 24.611/0.7609 & -/-\\
    
    T-DSSCAM & 316.29k& 26.7915/0.8104 & 26.5976/0.8154 & 29.4447/0.8418 & 23.673/0.7391 &	24.627/0.7612 & 0.016/0.0003\\
    
    T-NoSCA & 423.02k & 26.7377/0.8087 & 26.5269/0.8131	&29.3357/0.8402	& 23.6462/0.7364 & 24.588/0.7587 & -0.023/0.0022\\
    
    T-NAFGCBlock-1 & 413.82k & 26.7666/0.810 & 26.5542/0.8143 &	29.311/0.8411 &	23.6815/0.7387&	24.620/0.7607 & 0.009/-0.0002\\
    
    T-edge & 460.67k & 26.8062/0.8107 & 26.6275/0.8159 & 29.4241/0.8421 & 23.6982/0.7386 & 24.650/0.7609 & 0.039/0\\
    \hline
  \end{tabular}
  \label{TABLE5}
\end{table*}

\subsubsection{NAFGCBlock}
\label{sec:NAFGCBlock}
As described in Section \ref{sec:NAFGCBlock_stc}, NAFGCBlock is modified from NAFBlock by removing SCA and replacing PW convolution with 3$\times$3 GConv. We first rationalize the removal of SCA by evidencing that SCA is unnecessary. We still take the NAFSSR-T model as the baseline, and remove the SCA modules from all the NAFBlocks in the NAFSSR-T model to generate a new model called T-NoSCA. Comparative experiments are conducted between NAFSSR-T and T-NoSCA on the same datasets. 
As can be seen from Table~\ref{TABLE5}, on most test sets the PSNR and SSIM values of T-NoSCA are only slightly lower than those of NAFSSR-T. These results suggest that SCA is not crucial to the SR performance. The reason may be because the channel numbers are not large (e.g., the numbers of channels in NAFSSR-T, -S, and -B are 48, 64, 96, respectively), making the channel attention mechanism inefficient. Table~\ref{TABLE5} also shows that the parameter number of T-NoSCA is 91.8\% of that of NAFSSR-T. It is therefore demonstrated that removing SCA can reduce the model complexity while sacrificing little performance.

Having rationalized the removal of SCA, we proceed to investigate the effectiveness of NAFGCBlock which not only removes SCA but also replaces PW convolution with 3$\times$3 GConv. Specifically, as described in Section 3.1.3, we construct two types of NAFGCBlock: NAFGCBlock-1, for pursuing higher performance, and NAFGCBlock-2, for pursuing a very small number of parameters. We here only compare the performance between NAFGCBlock-1 and NAFBlock, because NAFGCBlock-2 is always used together with the recursive connection, making the performance comparison unfair. Along this line, T-NAFGCBlock-1 is constructed by replacing all the NAFBlocks in NAFSSR-T with NAFGCBlock-1. As shown in Table~\ref{TABLE5}, T-NAFGCBlock-1 exhibits not only higher PSNR/SSIM values but also fewer parameters, demonstrating that NAFGCBlock is a simpler yet more powerful module than NAFBlock for intra-view information extraction.

\subsubsection{Edge detection}
To demonstrate the positive role of edge detection in the stero image SR, an trainable edge detection operator (see Section \ref{sec:Edge_method}) is added into the NAFSSR-T model, resulting in a model called T-edge. Both T-edge and NAFSSR-T are trained and tested using the same method. The results of the two models on all the test sets are shown in the Table~\ref{TABLE5}. It is seen that from NAFSSR-T to T-edge, the PSNR/SSIM values are improved with almost no increase in the number of model parameters. This verifies that the edge information is important for the quality of an SR image, and the use of the edge detection technique can help to improve the performance of an SR model.

\subsection{Comparison to state-of-the-arts methods}
We compare the 4 variants of NAFRSSR, including NAFRSSR-M, NAFRSSR-T, NAFRSSR-S and NAFRSSR-B (see Section \ref{sec:Overall_architecture} for their architectures), with a number of existing single- and stereo-image SR methods. The single image SR methods include VDSR\cite{VDSR}, EDSR\cite{EDSR}, RDN\cite{RDN}, and RCAN\cite{RCAN}, while the stereo image SR methods include StereoSR\cite{StereoSR}, PASSRnet\cite{PASSRnet}, SRRes+SAM\cite{SRRes+SAM}, IMSSRnet\cite{IMSSRnet}, iPASSR\cite{iPASSR} and SSRDE-FNet\cite{SSRDE-FNet}), NAFSSR\cite{NAFSSR}, SwinFIR\cite{SwinFIR}). All the SR models are trained on the same dataset, and their test results on various datasets are summarized in Table~\ref{TABLE6}.

\begin{table*}[htbp]
  \centering
  \caption{Quantitative results achieved by different SR models on the KITTI 2012\cite{KITTI2012}, KITTI 2015\cite{KITTI2015}, Middlebury\cite{Middlebury}, and Flickr1024\cite{flickr1024} datasets. P represents the number of parameters of the networks. Here, PSNR/SSIM values achieved on both the left images (i.e., Left) and a pair of stereo images (i.e., (Left + Right) /2) are reported. Ours results are highlighted in bold.}
  \begin{tabular}{c c c c c c c}
    \hline
    Method & Parameter & KITTI 2012 & KITTI 2015 & Middlebury & Flickr1024 & Average \\
     \hline
    VDSR\cite{VDSR} &	0.66M &	25.60/0.7722 &	25.32/0.7703&	27.69/0.7941&	22.46/0.6718&	23.391/0.7010 \\
   
    EDSR\cite{EDSR} &	38.9M &	26.35/0.8015 & 	26.04/0.8039 &	29.23/0.8397 &	23.46/0.7285&	24.341/0.7509 \\
   
    RDN\cite{RDN} &	22.0M &	26.32/0.8014 & 	26.04/0.8043 &	29.27/0.8404&	23.47/0.7295&	24.345/0.7517 \\
 
    RCAN\cite{RCAN} &	15.4M &	 26.44/0.8029 & 	26.22/0.8068 &	29.30/0.8397 &	23.48/0.7286&	24.391/0.7516 \\
  
    StereoSR\cite{StereoSR} &	1.42M &	24.53/0.7555 & 	24.21/0.7511 &	27.64/0.8022 &	21.70/0.6460&	22.569/0.6783 \\
   
    PASSRnet\cite{PASSRnet} &	1.42M &	26.34/0.7981 & 	26.08/0.8002 &	28.72/0.8236 &	23.31/0.7195&	24.221/0.7431 \\
   
    SRRes+SAM\cite{SRRes+SAM} &	1.73M &	26.44/0.8018 & 	26.22/0.8054 &	28.83/0.8290 &	23.27/0.7233&	24.227/0.7471 \\
    
    IMSSRnet\cite{IMSSRnet} &	6.89M &	26.43/- & 	26.20/- &	29.02/-	& -/- &	/  \\
  
    iPASSR\cite{iPASSR}&	1.42M&	26.56/0.8053& 	26.32/0.8084& 	29.16/0.8367& 	23.44/0.7287&	24.386/0.7521 \\
   
    SSRDE-FNet\cite{SSRDE-FNet}&	2.24M&	 26.70/0.8082& 	26.43/0.8118& 	29.38/0.8411& 	23.59/0.7352&	24.532/0.7576 \\
     \hline   
     &&&&&& \\
     
    NAFSSR-T\cite{NAFSSR}&	0.46M&	26.75/0.8099&	26.58/0.8154&	29.37/0.8404&	23.67/0.739&	24.614/0.7610 \\
   
    NAFSSR-S\cite{NAFSSR}&	1.56M&	26.93/0.8145& 	26.76/0.8203& 	29.72/0.8490& 	23.88/0.7468&	24.821/0.7680 \\
  
    NAFSSR-B\cite{NAFSSR}&	6.80M&	27.08/0.8181& 	26.91/0.8245& 	30.04/0.8568& 	24.07/0.7551&	25.005/0.7752 \\
    \\
   \hline
      &&&&&& \\
                      
    SwinFIR-T\cite{SwinFIR}&	0.89M&	26.68/0.8081& 	26.51/0.8135& 	29.48/0.8426& 	23.73/0.7400&	24.643/0.7613 \\
   
    SwinFIR\cite{SwinFIR} &	13.99M&	26.92/0.8148&	 26.74/0.8206& 	30.14/0.8582& 	24.14/0.7560&	25.016/0.7750 \\
     
    &&&&&& \\
    \hline
     &&&&&& \\
    \textbf{NAFRSSR-M}&	0.28M&	26.80/0.8113&	26.57/0.8156&	29.40/0.8429&	23.72/0.7403&	\textbf{24.657/0.7622} \\
   
    \textbf{NAFRSSR-T}&	0.34M&	26.85/0.8125&	26.65/0.8179&	29.48/0.8446&	23.78/0.7431&	\textbf{24.718/0.7648} \\
    
    \textbf{NAFRSSR-S}&	1.35M&	26.98/0.8167&	26.78/0.8220&	29.82/0.8517&	23.96/0.7514&	\textbf{24.891/0.7719 }\\
  
   \textbf{NAFRSSR-B}&	5.78M&	27.07/0.8193&	26.86/0.8248&	30.04/0.8588&	24.12/0.7585&	\textbf{25.033/0.7779} \\
     &&&&&& \\
     \hline

  \end{tabular}
  \label{TABLE6}
\end{table*}

\begin{figure}[!t]
  \centering
  \includegraphics[width=3.4in, keepaspectratio]{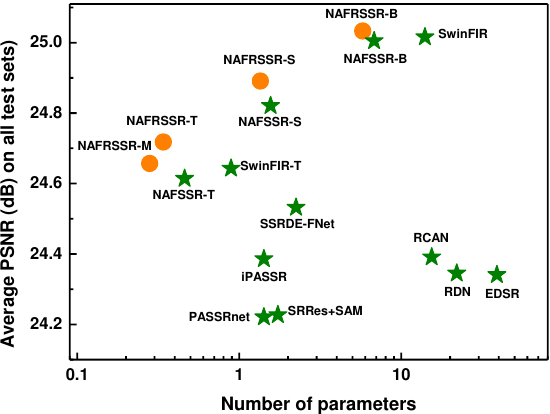}
  \caption{Parameter and PSNR comparison between our NAFRSSR models (-M, -T, -S, -B, depicted by orange solid circles) and the previous models (represented by green five-pointed stars).}
  \label{fig6}
\end{figure}

\subsubsection{Quantitative Evaluations}

As seen in Fig. \ref{fig6}, it is evident that our NAFRSSR models exhibit superior model performance (characterized by the average PSNR on all the test sets) with smaller model sizes than the previous state-of-the-art counterparts.

In particular, the number of parameters in our NAFRSSR-M model is only 0.28M, which is 60.8\% of NAFSSR-T and 31.5\% of SWinFIR-T. However, the NAFRSSR-M model achieves average PSNR/SSIM values of 24.657 dB/0.7622, which are 0.043 dB/0.0012 and 0.014 dB/0.0009 higher than those of NAFSSR-T and SWinFIR-T, respectively (see Table~\ref{TABLE6}).

In terms of our NAFRSSR-T model, its number of parameters is only 0.34M, which is still smaller than those of NAFSSR-T and SWinFIR-T. Meanwhile, NAFRSSR-T achieves average PSNR/SSIM values of 24.718 dB/0.7647, which are 0.101 dB/0.0038 and 0.072 dB/0.0035 higher than those of NAFSSR-T and SWinFIR-T, respectively (see Table~\ref{TABLE6}).

Our NAFRSSR-S model becomes slightly larger, which contains 1.35M parameters. Nevertheless, it is still 21.8\% smaller than NAFSSR-S in size. Besides, it achieves 0.07 dB/0.0036 higher average PSNR/SSIM values than NAFSSR-S (see Table~\ref{TABLE6}).

The NAFRSSR-B model is the largest one among our designed four NAFRSSR variants, and its number of parameters reaches 5.78M. Nevertheless, its model size is only 85\% of NAFSSR-B and 41.3\% of SwinFIR. Additionally, it achieves higher average PSNR/SSIM values (25.033 dB/0.7779) compared to NAFSSR-B (25.005 dB/0.7752) and SwinFIR\cite{SwinFIR} (25.016 dB/0.7750) (see Table~\ref{TABLE6}).

\subsubsection{Inference time}
We conduct a comparative study of the inference times of NAFRSSR and other advanced models, namely, NAFSSR\cite{NAFSSR}, iPASSR\cite{iPASSR}, PASSRnet\cite{PASSRnet}, SRRes+SAM\cite{SRRes+SAM}, RCAN\cite{RCAN}, and RDN\cite{RDN}. However, due to the unavailability of inference time data and source code, SwinFIR\cite{SwinFIR} is not included for comparison. The 128$\times$128 images are used as the test images, and the inference times are evaluated on GPU (NVIDIA GeForce RTX 3090). The hardware and software conditions are shown in Table~\ref{TABLE2} and \ref{TABLE3}. 

\begin{figure}[!t]
  \centering
  \includegraphics[width=3.4in, keepaspectratio]{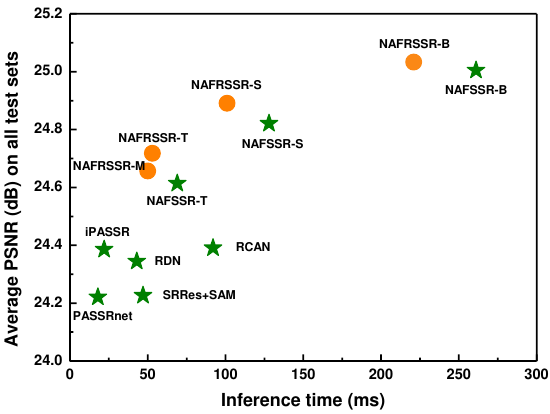}
  \caption{Inference time and PSNR comparison between our NAFRSSR (-M, -T, -S, -B, depicted by orange solid circles) and the previous model (represented by green five-pointed stars).}
  \label{fig7}
\end{figure}

It can be seen from Fig. \ref{fig7} that the inference times of our NAFRSSR-T, -S, and -B models are shorter than those of the NAFSSR-T, -S, and -B models, respectively. Also note that all the NAFRSSR models achieve higher average PSNR/SSIM values than the corresponding NAFSSR counterparts. This indicates that NAFRSSR is faster and more efficient compared with NAFSSR.
To realize a further speedup, we resort to the NAFRSSR-M model. As shown in Fig. \ref{fig7}, the inference time of NAFRSSR-M is as short as 50 ms. Therefore, to the best of our knowledge, NAFRSSR-M is the lightest and fastest stereo SR image model which can achieve average PSNR/SSIM values higher than 24.6 dB/0.76.

\begin{figure*}[!t]
  \centering
  \includegraphics[width=5.8in, keepaspectratio]{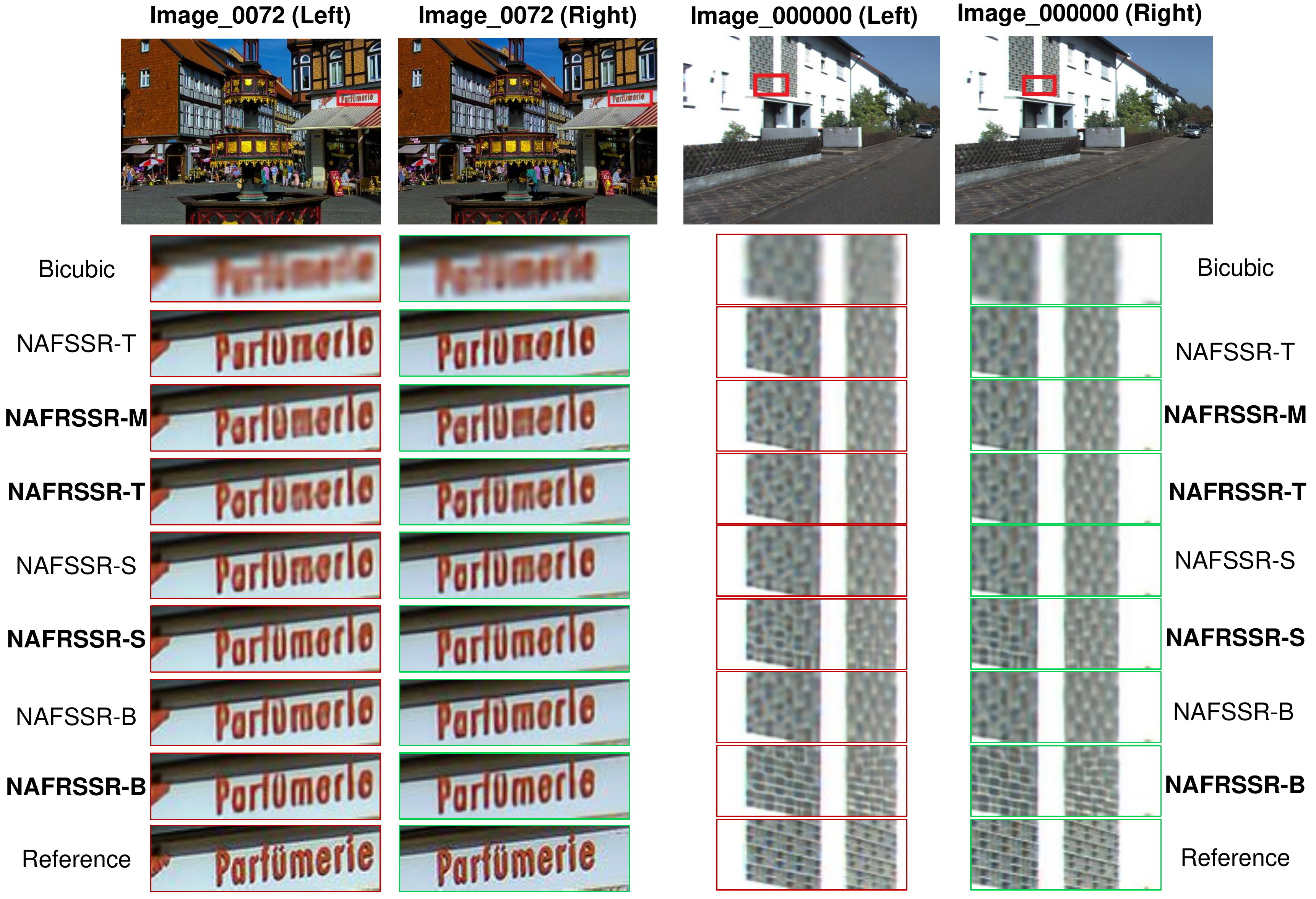}
  \caption{Visual results ($\times$4) achieved by bicubic, NAFSSR (-T,-S,-B) and NAFRSSR (-M, -T, -S, -B) models on the Flickr1024\cite{flickr1024} (left side) and KITTI 2012\cite{KITTI2012} (right side). The images with red and green borders represent the left and right views respectively.}
  \label{fig8}
\end{figure*}

\begin{figure*}[!t]
  \centering
  \includegraphics[width=6in, keepaspectratio]{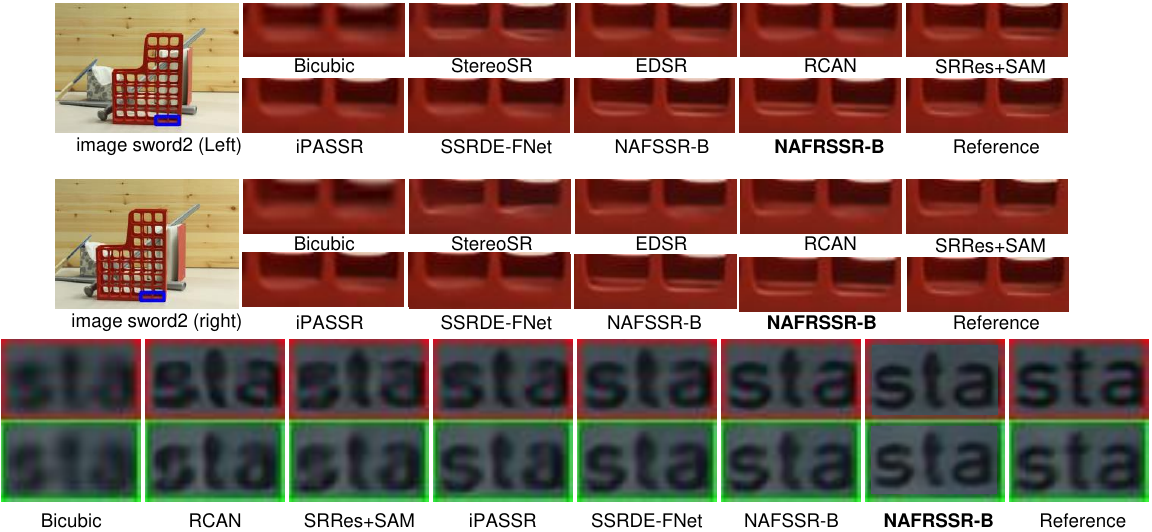}
  \caption{Visual results ($\times$4) achieved by different models on the Middlebury (top) and KITTI 2015\cite{KITTI2015} (bottom) dataset. In KITTI 2015, the images with red and green borders represent the left and right views respectively.}
  \label{fig9}
\end{figure*}

\subsubsection{Visual Comparison}
Fig. \ref{fig8} and \ref{fig9} shows the visual comparison results of $\times$4 SR images on the Flickr1024\cite{flickr1024}, KITTI 2012\cite{KITTI2012}, KITTI2015\cite{KITTI2015}, and Middlebury\cite{Middlebury} datasets generated by various SR models. As shown in Fig. \ref{fig8}, the visual effects achieved by our NAFRSSR-M/-T, -S, and -B models are superior to those of the previous NAFSSR-T, -S, and -B models, respectively. Furthermore, Fig. \ref{fig9} shows that the visual effect of our NAFRSSR-B model is better than those of other models. The SR images generated by NAFRSSR-B contains more detailed features and clearer edges than other SR images. In addition, the SR images reconstructed by other models still have some false texture features that do not exist in HR images. These results demonstrate that our NAFRSSR models have better visual effects.

\section{Conclusion}
In summary, we propose a lightweight recursive network, i.e., NAFRSSR, for efficient stereo image SR. NAFRSSR mainly consists of a stack of NAFGCBlocks (some are recursively connected) for intra-view feature extraction and DSSCAMs for cross-view feature fusion. NAFGCBlock and DSSCAM are modified from NAFBlock and SCAM of NAFSSR, respectively, through a lightweight design. In addition, we add a trainable edge detection operator to the final output layer of NAFRSSR to improve the quality of the reconstructed image. To show the superior performance of NAFRSSR, four variants of NAFRSSR with different sizes, i.e., NAFRSSR-M, NAFRSSR-T, NAFRSSR-S, and NAFRSSR-B are designed. They all exhibit fewer parameters, higher PSNR/SSIM, and faster speed than the previous state-of-the-art models, revealing their great application potential in stereo image SR.

\begin{IEEEbiography}[{\includegraphics[width=1in,height=1.25in,clip,keepaspectratio]{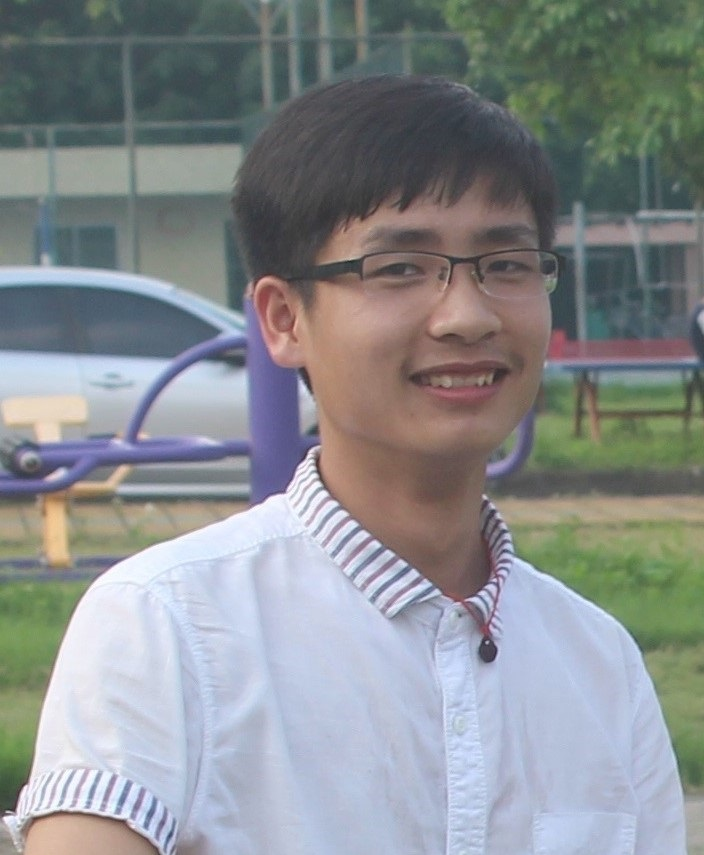}}]{Yihong Chen}
received the B.S. degree in electronic information science and technology from Jinan University, China, in 2020. He is currently pursuing a M.S. degree at South China Normal University. His research interests include memristive neural networks, computer vision, and natural language processing.
\vspace{-1ex}
\end{IEEEbiography}

\begin{IEEEbiography}[{\includegraphics[width=1in,height=1.25in,clip,keepaspectratio]{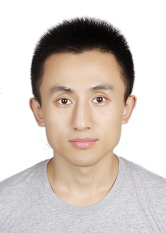}}]{Zhen Fan} received the Ph.D. degrees from National University of Singapore
(NUS) in 2015. He currently works as a Professor at South China Normal University, China. His research focuses on emerging nonvolatile memories, memristive neural networks, and machine learning.
\vspace{-1ex}
\end{IEEEbiography}

\begin{IEEEbiography}[{\includegraphics[width=1in,height=1.25in,clip,keepaspectratio]{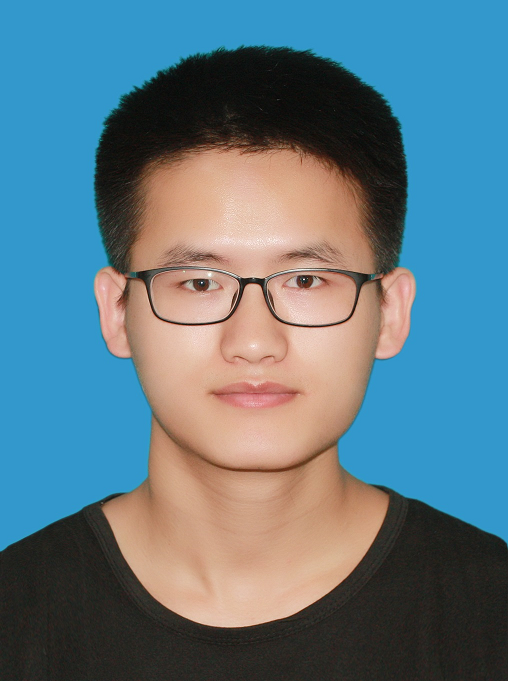}}]	{Shuai Dong} received the B.S. degree in electronic information engineering from Chongqing University, China, in 2018. He is currently studying for a M.S. degree at South China Normal University. His research interests include deep learning, computer vision and their hardware implementation using memristive neural networks.
\vspace{-1ex}
\end{IEEEbiography}

\begin{IEEEbiography}[{\includegraphics[width=1in,height=1.25in,clip,keepaspectratio]{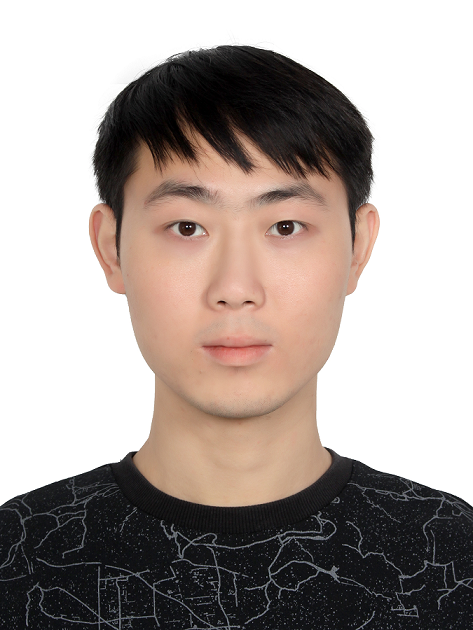}}]	{Zhiwei Chen} received the BS degree in Electronic lnformation Engineering from Nanjing University of Posts and Telecommunications, China in 2021. He is currently pursuing a M.S. degree at South China Normal University. His research interests include computer vision, memristive neural networks, and neuromorphic computing.
\vspace{-1ex}
\end{IEEEbiography}

\begin{IEEEbiography}[{\includegraphics[width=1in,height=1.25in,clip,keepaspectratio]{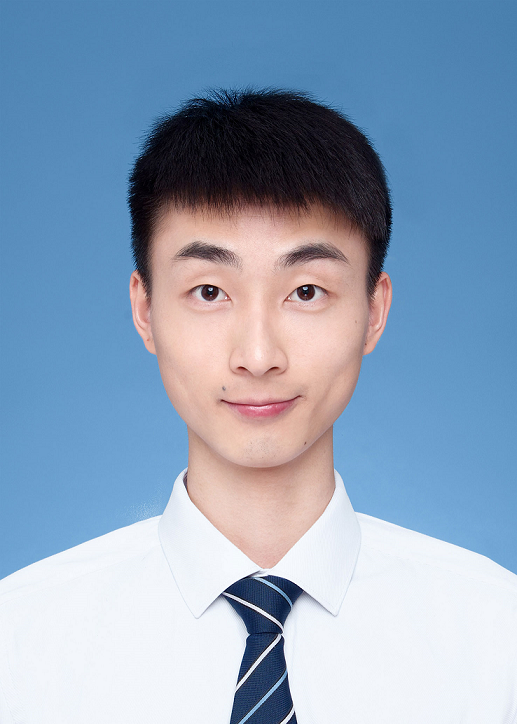}}]	{Wenjie Li} received the B.S. degree in material physics from South China Normal University, China, in 2019. He is currently pursuing a M.S. degree at South China Normal University. His research interests include neural networks, memristor and its application.
\vspace{-1ex}
\end{IEEEbiography}

\begin{IEEEbiography}[{\includegraphics[width=1in,height=1.25in,clip,keepaspectratio]{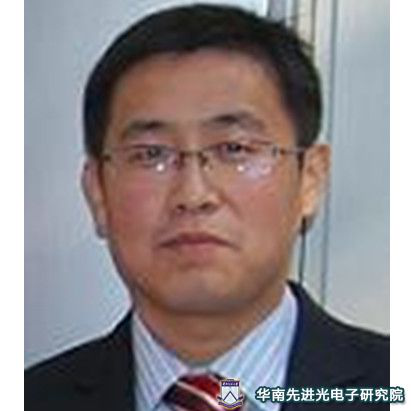}}]{Minghui Qin} received the Ph.D. degree in condensed matter physics from the Nanjing University, China, in 2010. He is currently a Professor at South China Normal University, China. His research focuses on magnetic materials and magnetic random access memory.
\vspace{-1ex}
\end{IEEEbiography}

\begin{IEEEbiography}[{\includegraphics[width=1in,height=1.25in,clip,keepaspectratio]{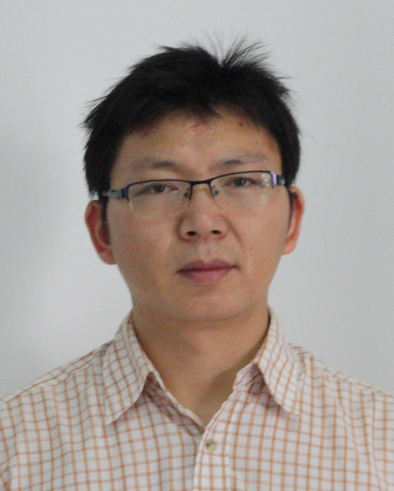}}]{Min Zeng}	
received the M.S. degree from School of Materials Science and Engineering, Jingdezhen Ceramic Institute, China, in 2004. He received his Ph. D degree from Department of Applied Physics, The Hong Kong Polytechnic University, Hong Kong, in 2010. He is currently a Professor at South China Normal University, China. His research topics are preparation, characterization and device design of multiferroic materials for memory and energy storage applications.
\vspace{-1ex}
\end{IEEEbiography}

\begin{IEEEbiography}[{\includegraphics[width=1in,height=1.25in,clip,keepaspectratio]{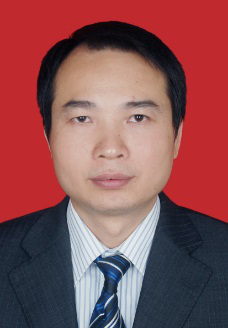}}]{Xubing Lu}
received the M.S. degree from Central South University, China, in 1999. He received his Ph. D degree from Nanjing University, China, in 2002. He is currently a Professor at South China Normal University, China. His research topics are high-K thin films, ferroelectric films and their device applications.
\vspace{-1ex}
\end{IEEEbiography}

\begin{IEEEbiography}[{\includegraphics[width=1in,height=1.25in,clip,keepaspectratio]{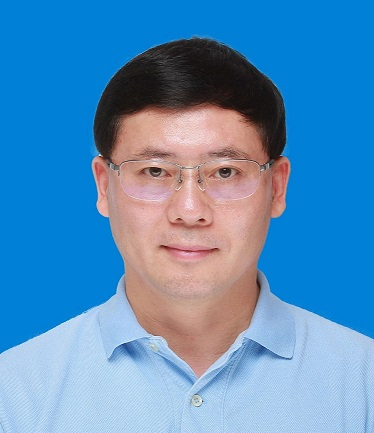}}]{Guofu Zhou}
received the M.S. degree in materials science from Chinese Academy of Sciences, China, in 1989. He received the first Ph. D degree in the same institute in 1991, and the second Ph. D degree in physics from Amsterdam University, Netherlands, in 1994. He is currently a Professor at South China Normal University, China. His research interests include emerging display technology, green optoelectronic materials and devices, and smart material technology.
\vspace{-1ex}
\end{IEEEbiography}

\begin{IEEEbiography}[{\includegraphics[width=1in,height=1.25in,clip,keepaspectratio]{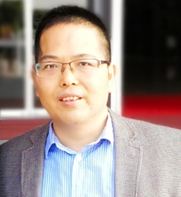}}]{Xingsen Gao}	
received the M.S. degree from Nanjing University, China, in 1999. He received his Ph. D degree in material science from National University of Singapore, Singapore, in 2004. He is currently a Professor at South China Normal University, China. His research interests include multi-functional scanning probe microscope testing, multi-functional ferroic thin films and nanostructures, and resistive switching materials and devices.
\vspace{-1ex}
\end{IEEEbiography}

\begin{IEEEbiography}[{\includegraphics[width=1in,height=1.25in,clip,keepaspectratio]{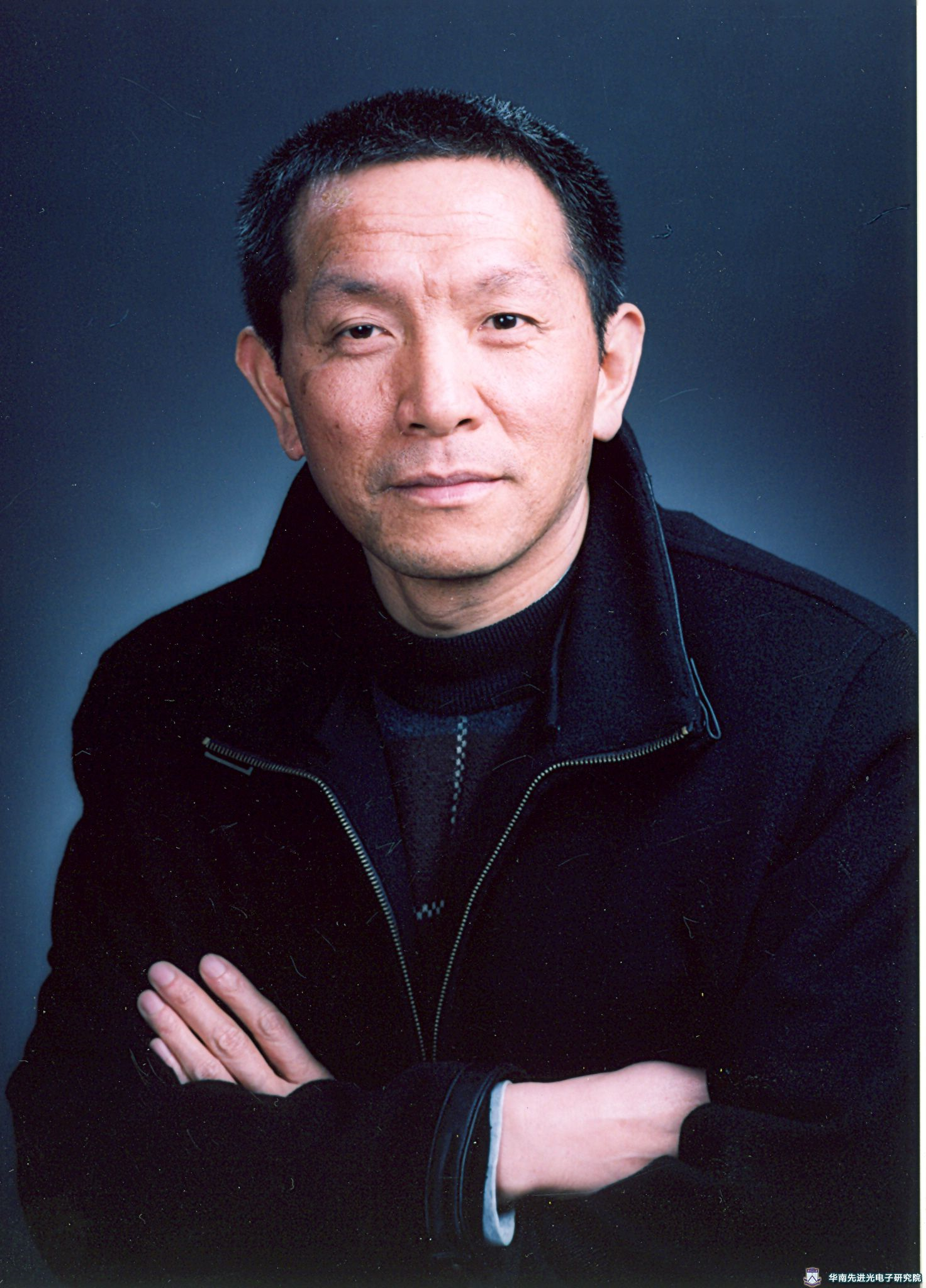}}]{Jun-Ming Liu}	
received the M.S. degree from Huazhong University of Science and Technology, China, in 1987. He received his Ph. D degree in material science from Northwestern Polytechnical University, China, in 1989. He is currently a Professor of School of Physics at Nanjing University. His research interests include experimental and theoretical research in correlated-electron physics, multiferroic and ferroelectric physics, statistical physics, and spintronic physics. 
\vspace{-1ex}
\end{IEEEbiography}


\end{document}